\documentclass[useAMS,usenatbib,usegraphicx]{mn2e}

\usepackage{amsmath}
\usepackage{url}

\def\fermi{{\it Fermi\/}}

\title[Mining for dark matter]{{\it Fermi's Sibyl}: Mining the gamma-ray 
sky for 
dark matter subhaloes}
\author[Mirabal et al.]{N. Mirabal$^{1,2}$\thanks{E-mail:
mirabal@gae.ucm.es}, V. Fr\'ias-Martinez$^{3}$, T. Hassan$^{2}$, and E. 
Fr\'ias-Martinez$^{3}$\\
$^{1}$Ram\'on y Cajal Fellow\\
$^{2}$ Dpto. de F\'isica At\'omica,
Molecular y Nuclear, Universidad Complutense de
Madrid, Spain\\
$^{3}$ Telefonica Research, Madrid, Spain\\
}

\begin{document}

\date{}

\pagerange{\pageref{firstpage}--\pageref{lastpage}} \pubyear{2012}

\maketitle

\label{firstpage}

\begin{abstract}
Dark matter annihilation signals coming from Galactic subhaloes  
may account for a small fraction of unassociated point 
sources detected in the Second {\it Fermi}-LAT catalogue (2FGL). 
To investigate this possibility, we present {\it Sibyl}, a Random 
Forest classifier that offers predictions on 
class memberships for unassociated {\it Fermi}-LAT sources at high Galactic
latitudes 
using gamma-ray features extracted from the
2FGL. {\it Sibyl} generates a large 
ensemble of classification trees that are trained
to vote on whether a particular object is
an active galactic
 nucleus (AGN) or a pulsar. 
After training on a list of 908 identified/associated 2FGL sources, 
{\it Sibyl} reaches individual accuracy rates of up to 97.7\% for AGNs and
96.5\% for pulsars. Predictions for the 269 unassociated 2FGL sources 
at $|b| \geq 10^\circ$ suggest that    
216 are potential AGNs and 16 are potential pulsars (with majority votes
greater than 70$\%$). The remaining 37 objects are inconclusive, 
but none is an extreme outlier.   
These results could guide future quests for dark matter Galactic 
subhaloes.
\end{abstract}


\section{Introduction}
The extraordinary success of the {\it Fermi} mission marks the
beginning of the golden age for gamma-ray astrophysics.
With 24 months of data, the Second {\it Fermi}-LAT catalogue (2FGL) lists
1873 sources  in the 100 MeV to 100 GeV energy range,
of which 886 are AGNs and 108 are pulsars.
While {\it Fermi}
has greatly mitigated issues inherent to source localisation in the gamma-ray
regime,
269 sources in the 2FGL
(15\% of the total)  remain without obvious counterparts 
at Galactic latitude $|b| \geq 10^\circ$.
Such failure to associate the entire {\it Fermi} catalogue
continues to fuel speculation about the existence of 
new types of gamma-ray source classes.

Probably the most intriguing potential sources of gamma ray emission are 
dark matter subhaloes \citep{diemand,springel}. 
Numerical cold dark matter (CDM) simulations suggest that galaxies like our 
own are
surrounded by a wealth of small dark matter 
subhaloes that survived 
structure formation \citep{klypin,moore}. 
Massive subhaloes ($M \geq 10^{7} M_{\odot}$)
would correspond to ``classical'' dwarf galaxies. Less massive ones  
would be optically elusive and might only be revealed
as gamma-ray point sources when weakly interacting massive 
particles
(WIMPs) annihilate to gamma rays \citep{kuhlen}. As a result, nearby 
dark matter subhaloes 
might be lurking among the unassociated {\it Fermi} sources at high Galactic latitudes. 
If found, an annihilation signal from 
Galactic subhaloes would clinch the first 
non-gravitational signature of dark 
matter.

The hunt for dark matter subhaloes in the {\it Fermi} catalogue 
is currently underway
\citep{nieto,belikov,
zechlin,acker12}. Most approaches
involve the hypothesised sharp spectral cut-off or step expected
at the WIMP mass \citep{berg}. Assuming that the WIMP mass
falls between 100 MeV and 50 GeV, a dark matter subhalo could
be detectable in the {\it Fermi} MeV-GeV band, but would
disappear in the GeV-TeV band, effectively creating a
TeV dropout.

Here we investigate the possibility of identifying dark matter 
subhalo candidates using supervised machine learning 
algorithms. Rather than starting 
with an {\it ad hoc} theoretical dark matter spectrum we
would like to exploit pattern recognition 
of known gamma-ray features in associated 
sources and use this information to locate 
outliers that might constitute
novel emitters. Machine learning algorithms have already 
been used to study the First
{\it Fermi} LAT Catalogue (1FGL). 
For example, \citet{class} investigated classification trees and 
logistic regression to predict classes of unassociated sources in
the 1FGL based on 
a set of gamma-ray
features. K-means clustering
was also applied to help distinguish individual counterparts 
within {\it Fermi} error contours  \citep{mirabal}.  

With an additional year
of collected {\it Fermi} data, the gamma-ray features reported 
in the 2FGL have improved substantially.  
In this paper we train the Random Forest classifier \citep{breiman} on 
identified/associated {\it Fermi} objects and build a set of decision trees 
that provide
predictions for high-latitude unassociated {\it Fermi} objects in the 2FGL.  
The paper is organised as follows. In section \ref{datasets} we
describe the datasets and the Random Forest algorithm. 
Section \ref{appli} describes the performance of the classifier
on unassociated {\it Fermi}
sources. Section \ref{subhalos} details the search for potential 
outliers. Finally, we provide our conclusions and discuss future work 
in section \ref{discuss}.

\begin{figure}
\hfil
\includegraphics[width=3.1in,height=2.82in,angle=0.]{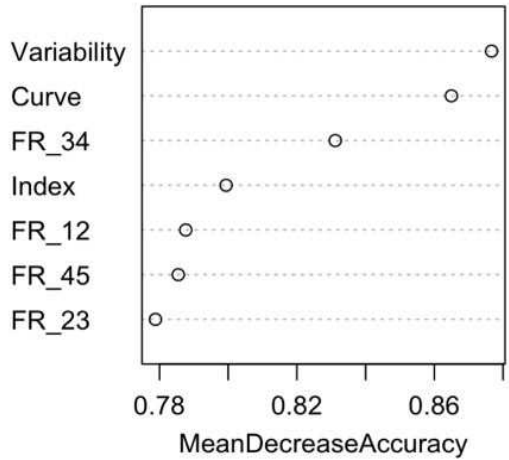}
\hfil
\caption{Gamma-ray features ranked in order of importance.
MeanDecreaseAccuracy measures  
the difference between accuracy rates before and after 
permutation of individual features averaged over all trees. 
Higher percentages indicate more
importance. 
}
\label{figure1}
\end{figure}

\section{Datasets  and Random Forests}
\label{datasets}
Random Forest is an ensemble classifier that grows a large 
forest of classification
trees \citep{breiman}. Decision trees are classification tools that have 
a tree structure, where each split is based on
    the information gained considering the elements of the feature space 
\citep{quinlan}.
To classify a new object, each tree in
the forest votes on the class. The proportion
of votes $P$ that agree on a decision
provides a measure of the accuracy of the classification.
Random Forest then makes a prediction based on the
majority of votes ($P>0.5$).
Random Forest 
also computes proximities
between pairs of objects and produces 
scaling coordinates (1st and 2nd) that can be used to
visualise datasets easily\citep{cox}.
In addition, a number of comparisons have 
shown that Random Forest is unexcelled in 
accuracy among current classifiers \citep{svetnik03,qi,frias,richards}.
The analysis presented here uses the
R randomForest package \citep{liaw}.

\begin{figure}
\hfil
\includegraphics[width=3.1in,angle=0.]{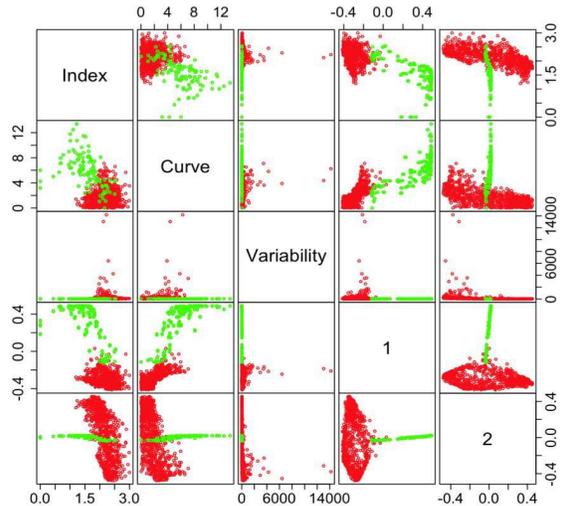}
\hfil
\caption{Properties of {\it Fermi} features plotted against each other.
Top features include index, curve, variability, and
1st and 2nd scaling coordinates (1 and 2 respectively) generated by 
{\it Sibyl}.
Two distinct classes are clear: AGNs (red)
and pulsars (green).}
\label{figure2}
\end{figure}

\begin{table*}
\caption{Predictions and voting percentages for unassociated {\it Fermi} sources in the 2FGL, ordered by RA}
\begin{tabular}{l c c l}
\hline
Source             & P$_{AGN}$ & P$_{Pulsar}$ & Prediction \\
\hline
2FGL J0004.2+2208 &  0.974 & 0.026 & AGN\\
2FGL J0014.3--0509 & 0.992 & 0.008 & AGN\\
2FGL J0031.0+0724 &  0.946 & 0.054 & AGN\\
2FGL J0032.7--5521 & 0.998 & 0.002 & AGN\\
2FGL J0039.1+4331 &  0.776 & 0.224 & AGN\\
2FGL J0048.8--6347 & 0.922 & 0.078 & AGN\\
2FGL J0102.2+0943 &  0.998 & 0.002 & AGN\\
2FGL J0103.8+1324 &  0.998 & 0.002 & AGN\\
2FGL J0106.5+4854 &  0.406 & 0.594 & --\\
2FGL J0116.6--6153 & 0.992 & 0.008 & AGN\\
2FGL J0124.6--2322 & 0.998 & 0.002 & AGN\\
2FGL J0129.4+2618  & 0.820 & 0.180 & AGN\\
2FGL J0133.4--4408  & 0.968 & 0.032 & AGN\\
2FGL J0143.6--5844  & 1.000 & 0.000 & AGN\\
2FGL J0158.4+0107  & 0.990 & 0.010 & AGN\\
\hline
\end{tabular}
\\Note: The complete list  of predictions is available
at \url{http://www.gae.ucm.es/~mirabal/sibyl.html}
\label{table1}
\end{table*}

For our dataset, we collected the complete \fermi\ LAT 
2FGL catalogue that consists of 1873 sources  (100 MeV-100 GeV) 
of which 1300 are firmly identified/associated and 573 are
unassociated sources
\citep{2lat,2lac}. In total,  
we consider a list that includes 
800 labelled AGNs (BL Lacs and flat-spectrum radio quasars only) 
and 108 pulsars. There are additional gamma-ray classes in
the 2FGL, but AGNs and pulsars are the largest and most common at 
$|b| \geq 10^\circ$. Thereby we simply consider a bimodality of classes.
Novelty detection will be discussed later on.  
For each of the 908 sources a total 
of 68 features are reported in the 2FGL. Features include Galactic latitude,
Galactic longitude, spectral 
index (Index), 
variability, curvature index (Curve), and fluxes in five bands. In addition, 
we 
generate four derived features defined by flux 
ratios $FR_{ij} = Flux_i/ Flux_j$ between consecutive 
bands for 0.1--0.3 GeV (Band 1), 0.3--1 GeV (Band 2), 1--3 GeV (Band 3), 
3--10 GeV (Band 4), 
and
10--100 GeV (Band 5) comparable to the features  
first introduced by \citet{class}. 

To avoid working with too many features that could generate
noise in the classifier, we first identify the subset
of features  
that best discriminates what constitutes an AGN or a pulsar. 
For that purpose, we compute the relevance of each feature towards
the target class, rank them by importance, and apply the classifier to
a subset of the most relevant ones.
Specifically,
we use the measure of importance 
-- MeanDecreasedAccuracy -- implemented within randomForest 
\citep{breiman,svetnik}. Initially,  
the accuracy rate is computed for each tree as the Random Forest 
is constructed.
The value of a particular 
feature is then
permuted across all the objects while other features are left unchanged. 
and the accuracy rate is recorded again. The MeanDecreaseAccuracy 
is the overall percentage decrease in accuracy rate averaged over 
all trees. If the feature is 
important,
there should be a greater decrease in the accuracy rate
compared to the initial one. Figure \ref{figure1} shows
the top most important features ranked by importance.
We found that the 
features that most clearly differentiate AGN and pulsar classes include: 
Index, Curve, Variability, 
and Flux Ratios ($FR_{12}$, $FR_{23}$, $FR_{34}$, and $FR_{45}$).
This selection is in general agreement with  \citet{class} who chose 
similar features for supervised classification of the 1FGL. 
Additional features showed considerably smaller values in
their importance (MeanDecreaseAccuracy) and are thus not considered 
in the analysis.

In order to construct and train {\it Sibyl}\footnote{In ancient Greece, a sibyl
was a person or agency considered to be a source of predictions or
oracles.}, 
we start with the 800 AGNs and 108 pulsars. 
However, given the highly imbalanced nature of the sets, we
replicate the pulsar sample to
attain a closer size as the AGN class
\citep{ling,chawla}. Practically, the content of the datasets
have not changed but the
replication mechanism adds weight to the minority sample and achieves
improved performance in the classifier. 

After matching the AGNs and
pulsar datasets, we generate 100 alternate training  
and testing sets built from randomly selected objects (2/3 and 1/3 of 
the sample respectively). We next 
produce Random Forest models with 500 trees for each training set. 
For validation, 
individual performance is evaluated at each of the 100 testing sets. 
Accuracy rates are computed directly by comparing the
class predicted by {\it Sibyl} with the true class for each object in the
testing sets. On average, {\it Sibyl} achieves an accuracy rate of 97.1\%
based on majority voting (97.7\% for AGNs and 96.5\% for pulsars).
Inclusion of absolute Galactic
latitude $|b|$ in the classifier 
lowered AGN and pulsar accuracy rates slightly 
to 97.4\%  to 95.5\% respectively. Since pulsars tend to be situated along
the Galactic plane and AGN are more numerous at high Galactic latitude, it
is possible that using
Galactic latitude as a feature 
could introduce a tiny bias against
AGN near the Galactic plane and pulsars away from it \citep{class}.
Generally, most of the misclassifications occur when 
less than 70\% of the individual trees ($P<0.7$) agree on a classification.
Figure \ref{figure2} displays the
outstanding separation between AGNs and pulsars, 
which explains the high accuracy rates obtained by {\it Sibyl}.

\section{Application to unassociated sources}
\label{appli}
The designation of 2FGL sources usually falls into three categories: 
identified, associated, and unassociated. 
A firm identification of a gamma-ray source
can only be established through
contemporaneous temporal variability, similar
spatial morphology, or equivalent pulsation at other wavelengths.
An association only requires positional correspondence of a possible
counterpart with a 2FGL source. Unassociated sources 
lack a formal counterpart at other wavelengths.  

Here, we consider a fourth category to designate 2FGL sources: 
``prediction''. Our objective is twofold: to predict the class of
high-latitude unassociated {\it Fermi} objects in the 2FGL;
and to produce a list of outliers that could be explored as potential
dark matter subhaloes. 
 For each of the  
269 unassociated {\it Fermi} sources at $|b| \geq 10^\circ$,
{\it Sibyl} provides a prediction that the object is
an AGN (P$_{AGN}$)  or a pulsar (P$_{Pulsar}$) based on 
individual votes polled from all trees in the decision forest. 

Since we want to isolate outliers 
that might constitute dark matter subhalo candidates, we
only accept {\it Sibyl} predictions 
whenever P $>0.7$ \emph{i.e.,} at least 70$\%$ of the trees
agree on the final decision. Otherwise, the object
remains without a prediction.
Such threshold value is set based on the
results explained in Section \ref{datasets}.
In total, {\it Sibyl} predicts
216 objects to be AGN and 16 to be pulsars.
The resulting predictions and percentages of voting agreements
are listed in Table \ref{table1}.
Finally, the remaining 37 objects left without a firm prediction are 
the focus of our outlier study in the next section. It is 
important to note that under some specific circumstances,
dark matter subhaloes could mimic the spectral properties of certain pulsars 
\citep{baltz,zechlin}, we discuss this possibility further in
\ref{discuss}.

\section{Search for dark matter subhaloes in the 2FGL}
\label{subhalos}
\begin{figure}
\hfil
\includegraphics[width=3.1in,angle=0.]{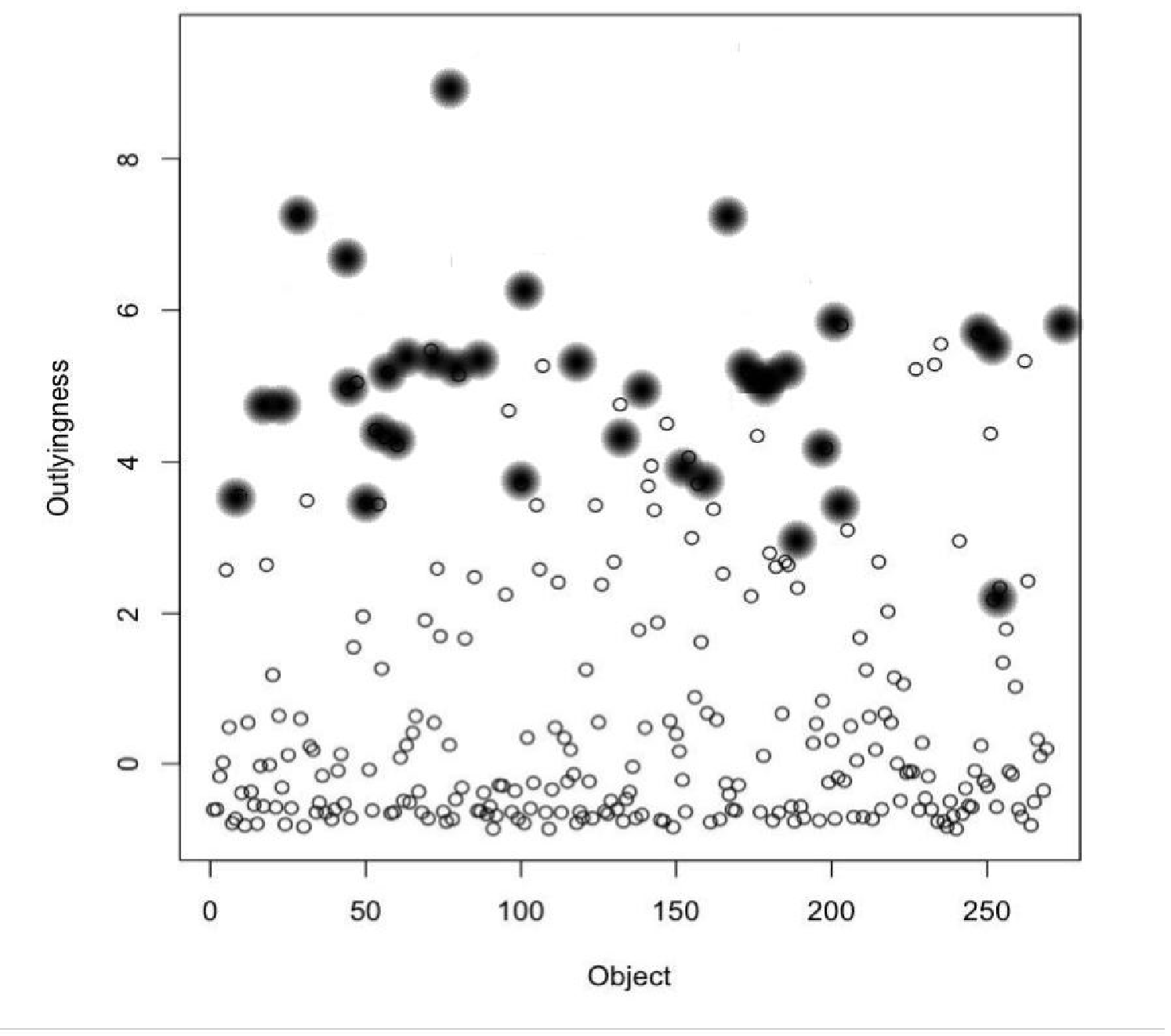}
\hfil
\caption{Distribution of outlyingness for the 37 objects
without firm predictions (shaded circles) and the
232 predicted by {\it Sibyl} (open circles).
The top outliers are summarised in Table \ref{table2}.
}
\label{figure3}
\end{figure}

In order to better understand the nature of the remaining 37 
objects we want compute their outlyingness, which
is a measure of how far away an object is from its closest class.
Apart for predicting an object's class, Random Forest computes 
the proximity of each predicted {\it Fermi} object $n$ 
to every element $k$ within each class $\sum_{\epsilon class} prox(n,k)$.
Formally, each individual proximity $prox(n,k)$ is computed as
the fraction of trees in which elements $n$ and ${k}$
fall in the same terminal node \citep{breiman,liaw}.
The outlyingness of an element $n$
is calculated as the reciprocal sum
of the squared proximities to all objects within its class.
This outlying measure is normalised
by subtracting the median and dividing by the                                  
absolute deviation from the median \citep{liaw}. 
Larger outlyingness values are common in objects that are 
extremely different
from the average, which could correspond to dark matter subhaloes. 
Figure \ref{figure3} shows the distribution of outlyingness 
for the 37 objects without firm predictions.
For comparison, we also plot the outlyingness for the
remaining 232 objects that were predicted by {\it Sibyl}
in the previous section.
Additionally, Table \ref{table2} lists the five objects with the largest 
outlyingness.

\begin{table}
\caption{Top outliers among high-latitude unassociated sources in the 2FGL}
\begin{tabular}{l c c}
\hline
Source             & P$_{AGN}$                                & Outlyingness\\
\hline
2FGL J0953.6--1504 &  0.658 & 9.0\\
2FGL J0418.9+6636 &  0.574 & 7.2\\
2FGL J1710.0--0323 & 0.500  & 7.1\\
2FGL J0533.9+6759 &  0.336 & 6.6\\
2FGL J0336.0+7504 & 0.476  & 6.2\\
\hline
\end{tabular}
\label{table2}
\end{table}

Given that outlyingness values much 
greater than 10 usually indicate novel cases \citep{breiman},
there is 
no strong indication of novelties (significant outliers) among the 37
objects without firm predictions. 
We find that the top five outliers have an 
average flux of $1.1 \times 10^{-9}$ ph cm$^{-2}$ s$^{-1}$ (1--100 GeV).  
Unassociated source fluxes at high latitudes range
from $7.7 \times 10^{-9}$ to $1.1 \times 10^{-10}$ ph 
cm$^{-2}$ s$^{-1}$. Thus, they are not necessarily the faintest sources in
the dataset. On the other hand, the mean photon index of sources 
in Table \ref{table2} 
is $2.2 \pm 0.3$, while photon indices in the unassociated
sample range from 1.1 to about 3.0. Inspection of individual
features in this manner yields limited insight into what makes these outliers 
stand out from the rest of the sample. As mentioned before, the exploration of
the entire feature space is precisely 
where the supervised learning algorithm 
excels. Unfortunately, {\it Sibyl} 
cannot assess by itself whether the outlyingness is due to 
an anomaly in the data taking process, a simple variation within known 
{\it Fermi} classes,
or a true novel source class such as dark 
matter annihilation in Galactic subhaloes.

\section{Discussion and Conclusions}
\label{discuss}
We have presented the outcome of the Random Forest predictor
{\it Sibyl}. 
The results show that machine learning
algorithms provide a reasonable route not only to predict unassociated
AGNs/pulsars in the 2FGL, but also to produce a list of sources
with unusual features that could be explored as
potential dark matter subhalo candidates.
After training on 908 identified/associated {\it Fermi} objects, 
{\it Sibyl} has been applied  
to predict the class of unassociated {\it Fermi} sources in the 2FGL.
Out of 269 unassociated sources at high latitudes, we have 
found that 216 are AGN candidates
and 16 are considered potential pulsars with 
prediction accuracy rates greater than 96.5$\%$.
{\it Sibyl} has also produced a list of 37 outlier objects; however, 
none of these exhibits significant outlyingness that
can be directly connected 
to new gamma-ray classes (including dark matter subhaloes) at this point.
We emphasise again that our results are strict predictions based on 
pattern recognition and thus a rigourous source identification process will
have to localise actual counterparts at other wavelengths. 

The results leave some room, albeit very small, to 
accommodate dark matter subhaloes or alternative source classes in the 2FGL.
These pockets could be targeted to exhaust all 
possibilities. Looking forward, 
zooming in on a reduced group of sources might be a wise observational 
strategy. For obvious reasons, the set of objects 
with the largest outlyingness could be a reasonable place 
to conduct a dedicated 
survey.  If dark matter consists of particles with a mass 
below 60 GeV \citep{hooper}, 
dark matter subhaloes might also be camouflaging among
the ranks of predicted pulsars as their spectral
signature could be similar to the pronounced 
spectral cut-off predicted predicted by certain 
dark matter models. However, a number of 
these sources could be old radio-quiet pulsars which 
will complicate the search for a counterpart \citep{kerr}.

There are a number of issues that need further exploration.  
For instance, the predictions are heavily dependent on the robustness of the
spectral parameters listed in the 2FGL. Most machine learning algorithms 
lack a proper treatment of uncertainties in each of the features considered 
\citep{carroll,morgan}. Inclusion of uncertainties as individual features in 
{\it Sibyl} did not
yield improved performances in our predictions. 
With additional years of flight, 
{\it Fermi}  will likely keep  
improving the accuracy of the gamma-ray features. However,
attempts should be made to account for feature errors properly. 
In a forthcoming paper, we plan to explore a more refined 
breakdown into further {\it Fermi} subclasses \citep{hassan}. There 
are at least four AGN subclasses 
in the 
2FGL comprising BL Lacs, flat-spectrum radio quasars, misaligned AGNs, and 
Seyfert 
galaxies \citep{2lac}. Pulsars can be further partitioned 
into radio-loud, radio-quiet, and millisecond
pulsars \citep{pulsars}.

Ultimately, the main reason that a large {\it Fermi} fraction remains 
unassociated to begin with 
has to do with the quality of localisations in the gamma-ray band. 
At faint flux levels, it becomes ever more difficult
to associate a {\it Fermi} source with a particular counterpart.
The best association procedures rely on positional
coincidences and correlations
with flat-spectrum radio sources \citep{2lac}.
None the less, considering the results presented here
and the scatter in gamma-ray flux
it seems likely that many of the unassociated sources at high latitude
are AGNs or mid-latitude pulsars
with somewhat fainter radio fluxes than their brighter cousins.

Without a major breakthrough in localisations,
the actual counterparts of most unassociated {\it Fermi} objects 
will be difficult to pinpoint in the short term. Machine learning algorithms 
can help narrow the options.       
Eventually, we will see significant improvement in localisations, 
particularly for Galactic sources, 
courtesy of the future Cherenkov Telescope Array (CTA) that 
will achieve enhanced angular resolution above 25 GeV \citep{cta}.

\section*{Acknowledgments}
N.M. acknowledges support from the Spanish government 
through a Ram\'on y Cajal fellowship and the 
Consolider-Ingenio 2010 Programme under grant MultiDark CSD2009-00064. 
We thank Pablo Saz Parkinson 
for helpful conversations. We also thank the referee for useful 
suggestions and comments on the manuscript.

\label{lastpage}
\end{document}